\documentclass[aps,prl,twocolumn,groupedaddress,showpacs]{revtex4}

\usepackage{graphicx}
\usepackage{amsmath}

\begin{document}

\title{A Bose-Einstein condensate bouncing off a rough mirror}

\author{H\'el\`ene Perrin}
\author{Yves Colombe}
\altaffiliation[Present address: ]{Sektion Physik der Ludwig-Maximilians Universit\"at, Schellingstrasse 4, D-80799 M\"unchen, Germany.}
\author{Brigitte Mercier}
\altaffiliation[Present address: ]{Laboratoire d'Optique Appliqu\'ee, Centre de l'Yvette, F-91761 Palaiseau Cedex, France.}
\author{Vincent Lorent}
\affiliation{Laboratoire de Physique des Lasers, UMR 7538 du CNRS, Institut Galil\'ee, Universit\'e Paris-Nord, Avenue J.-B. Cl\'ement, F-93430 Villetaneuse, France}
\author{Carsten Henkel}
\affiliation{Institute of Physics, Potsdam University,\\
Am Neuen Palais 10, 14469 Potsdam, Germany}
\date{\today}

\begin{abstract} We present experimental results and theoretical analysis of the diffuse reflection of a Bose-Einstein condensate from a rough mirror. The mirror is produced by a blue-detuned evanescent wave supported by a dielectric substrate. The results are carefully analysed via a comparison with a numerical simulation. The scattering is clearly anisotropic, more pronounced in the direction of the evanescent wave surface propagation, as predicted theoretically.
\end{abstract}

\maketitle

\section{Introduction}

Since the first experimental realisation of a Bose-Einstein condensation on a micro-chip~\cite{Reichel01}, the study of the interactions between ultra cold atoms and surfaces is of crucial interest. For instance, the quality of the wires used in micro fabricated chips is directly linked to the fragmentation effects observed in trapped BEC near a metallic wire \cite{fragmentation}. Also, it has been shown than the thermal fluctuations of the current in a metallic surface could induce spin flip losses in an atomic cloud when the distance to the surface is smaller than $10~\mu$m typically \cite{spinflipsHenkel,spinflips}.

The use of dielectric surfaces and evanescent waves for producing strong confinement has not been so widely explored yet. It has the advandage of a strong suppression of the spin flip loss mechanism \cite{spinflipsHenkel}. With such a system, one can realize mirrors \cite{evmirrors}, diffraction gratings \cite{evdiffraction}, 2D traps \cite{2devtraps} or waveguides \cite{evwaveguides}. Experiments involving ultra cold atoms from a Bose-Einstein condensate at the vicinity of a dielectric surface started only recently, leading for instance to the realisation of a two dimensional BEC \cite{Grimm04} or to sensitive measurements of adsorbate electric polarization \cite{McGuirk04}.

In this paper, we present recent experimental results of Bose condensed atoms interacting with the light field of an evanescent wave above a dielectric slab. An optical waveguide on a dielectric substrate was designed for studying two dimensional geometries \cite{Colombe03}. When a blue detuned light is injected into this waveguide, it results in an atomic mirror for the condensed atoms released from a Ioffe Pritchard magnetic trap. We observe a strong scattering of the atomic cloud due to the surface roughness. Section~\ref{experiment} is devoted to the description of the experiment and the analyse of the results. Section~\ref{theory} presents the comparison with a theoretical model developped by Henkel \textit{et al.} \cite{Henkel97a}.

\section{Experiment}
\label{experiment}

\subsection{The evanescent wave mirror} 
\begin{figure}[ht]
\begin{center}
\includegraphics[height=20mm]{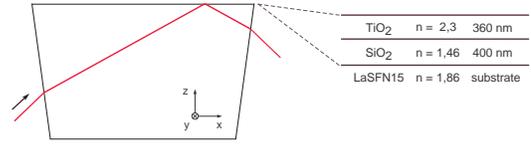}
\end{center}
\caption{Dielectric prism supporting the evanescent wave. The surface is coated by two layers of successively low and high refraction index to realize a wave guide and enhance the evanescent field. For each incident polarization, TE or TM, coupling is resonant for a given incident angle. The experiments were performed with TE polarization. One denotes $x$ as the propagation axis of the evanescent wave along the surface, $y$ as the other horizontal axis and $z$ as the vertical one.}
\label{prism}
\end{figure}

The atomic mirror is based on the short range repulsion of a blue detuned evanescent wave~\cite{evmirrors}. The evanescent wave is produced from a dielectric TiO$_2$ waveguide of index $n_g=2.3$ and thickness 360~nm, separated from a glass substrate of index 1.86 by a SiO$_2$ gap of index 1.5 and thickness 400~nm (figure~\ref{prism}). This surface coating was designed to enhance the evanescent field above the waveguide~\cite{Kaiser94}. The incident angle is fixed by the resonance condition for a TE polarization inside the waveguide. At the TiO$_2$ - vacuum interface, the resulting angle is $\theta_i = 46.1^{\circ}$ which leads to a decay length of the light field $\kappa^{-1}=93.8$~nm where $I = I_0 \, e^{-2 \kappa z}$. In the horizontal plane, the mode propagates along $x$ with a wave vector $K_{\rm ev} \mathbf{e_x}$ with $K_{\rm ev} = 1.66 k_L$, $k_L=2\pi/\lambda$ being the free space wave vector and $\lambda_L = 780$~nm the laser wavelength.

The light coupled into the waveguide is produced by a Gaussian shaped laser diode of power 40~mW detuned 1.5~GHz to the blue of the $5S_{1/2}, F=2 \longrightarrow 5P_{3/2}, F'=3$ D2 line of $^{87}$Rb. The Gaussian beam is elliptical and is focused at the prism surface to a $1/\sqrt{e}$ waist diameter of $100~\mu$m in the plane of incidence and $85~\mu$m along $y$. Here, one denotes $x$ as the propagation axis of the evanescent wave along the surface, $y$ as the other horizontal axis and $z$ as the vertical one (see figure~\ref{prism}). After projection onto the surface, the $x$ waist is $w_x = 220~\mu$m whereas the $y$ waist remains unchanged. The light intensity at the center of the spot at the surface is 210~W/cm$^2$. This value is deduced from a measurement of the light intensity threshold for the atomic reflection, taking into account the van der Waals attraction to the surface. Note that the number of photons scattered by each atom during reflection is about $n_{sp} \simeq 0.13$ only.

\subsection{Bouncing off the mirror}
Experimental details on the Bose-Einstein condensation set-up are given in Ref.~\cite{Colombe03}. $^{87}$Rb atoms are prepared and condensed in the $F=2, m_F=2$ state in a magnetic Ioffe-Pritchard type trap, elongated along $x$, with oscillation frequencies respectively $\omega_x/2 \pi = 21$~Hz along $x$ and $\omega_{\perp}/2 \pi = 220$~Hz in the radial directions. The center of the magnetic trap sits at $h_0=3.6$~mm above the evanescent mirror. This distance can be adjusted with of another pair of magnetic coils~\cite{ColombeCOLOQ}. About $N = 3\times10^5$ atoms cooled to below the condensation threshold are released from the magnetic trap. They experience a free fall of 27~ms before reaching the evanescent wave and being reflected. The laser beam creating the mirror is switched on for only 2.2~ms to avoid photon scattering during free fall and after reflection. This time is nevertheless sufficient to ensure a full reflection of the whole atomic cloud. An absorption picture is taken at the end of the experimental sequence, either before or after reflection.

From the pictures taken before reflection, we infer experimental parameters which will be useful for further analysis. One calibrates the pixel size by the center of mass motion under gravitation. We have also noticed an initial horizontal velocity $v_x = -30.67~$mm/s, due to a transient inhomogeneous magnetic field during the switching off procedure. From the balistic expansion, we get the fraction of condensed atoms ($N_0/N = 0.4$), the temperature of the thermal cloud $T=285$~nK, the Thomas-Fermi radius along $x$ $R_x=90~\mu$m and the velocity radius along $z$ $V_{\perp}=5.96$~mm/s. The velocity radius along $x$, $V_x$, is not measurable, being very small. However, following Ref.~\cite{Castin96}, it can be deduced from $V_{\perp}$ and the ratio of the oscillation frequencies in the magnetic trap: $V_x = \frac{\pi}{2} \, \frac{\omega_x}{\omega_{\perp}} V_{\perp} = 0.89$~mm/s.

\begin{figure}[ht]
\begin{center}
\includegraphics[width=40mm]{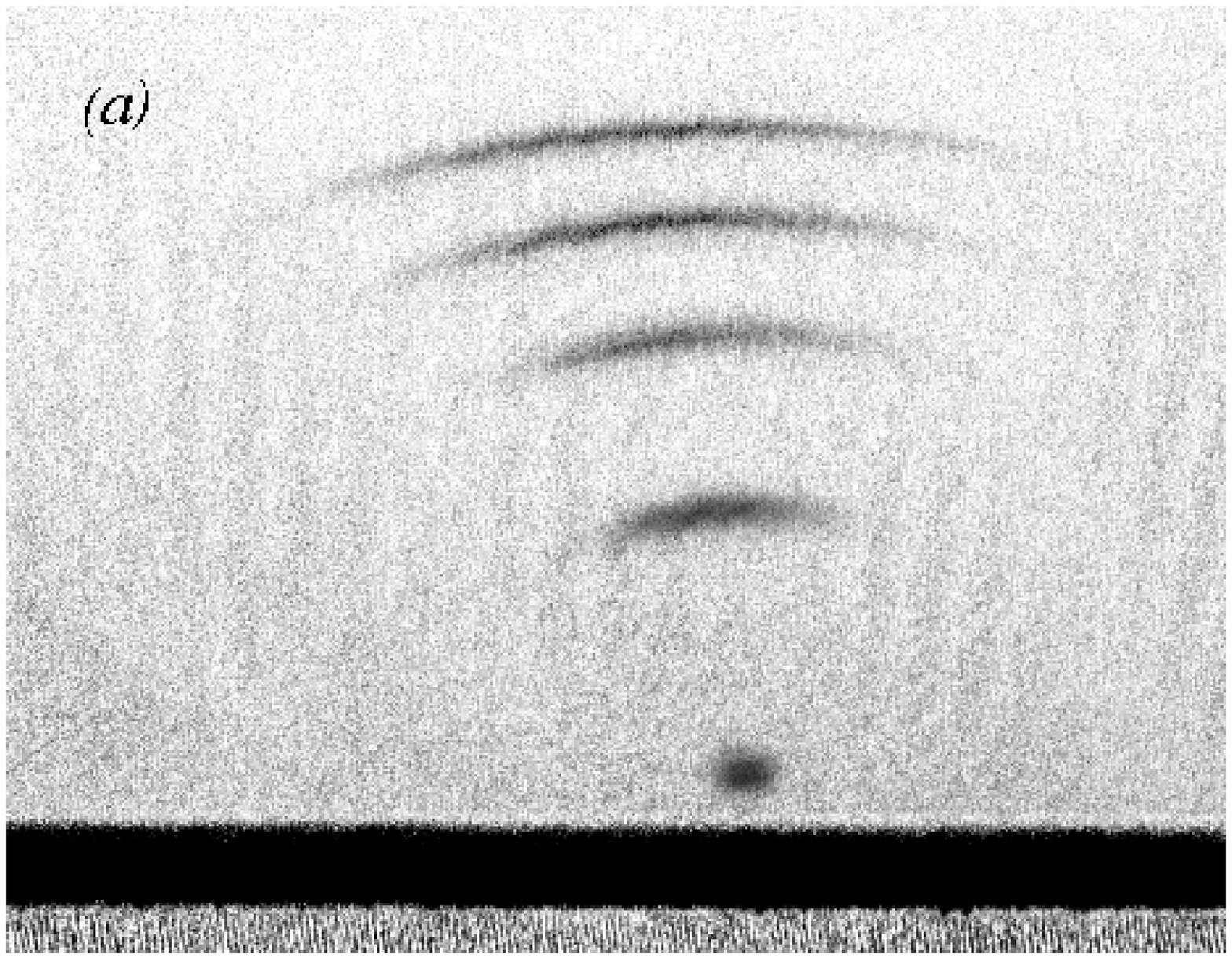} \quad \includegraphics[width=40mm]{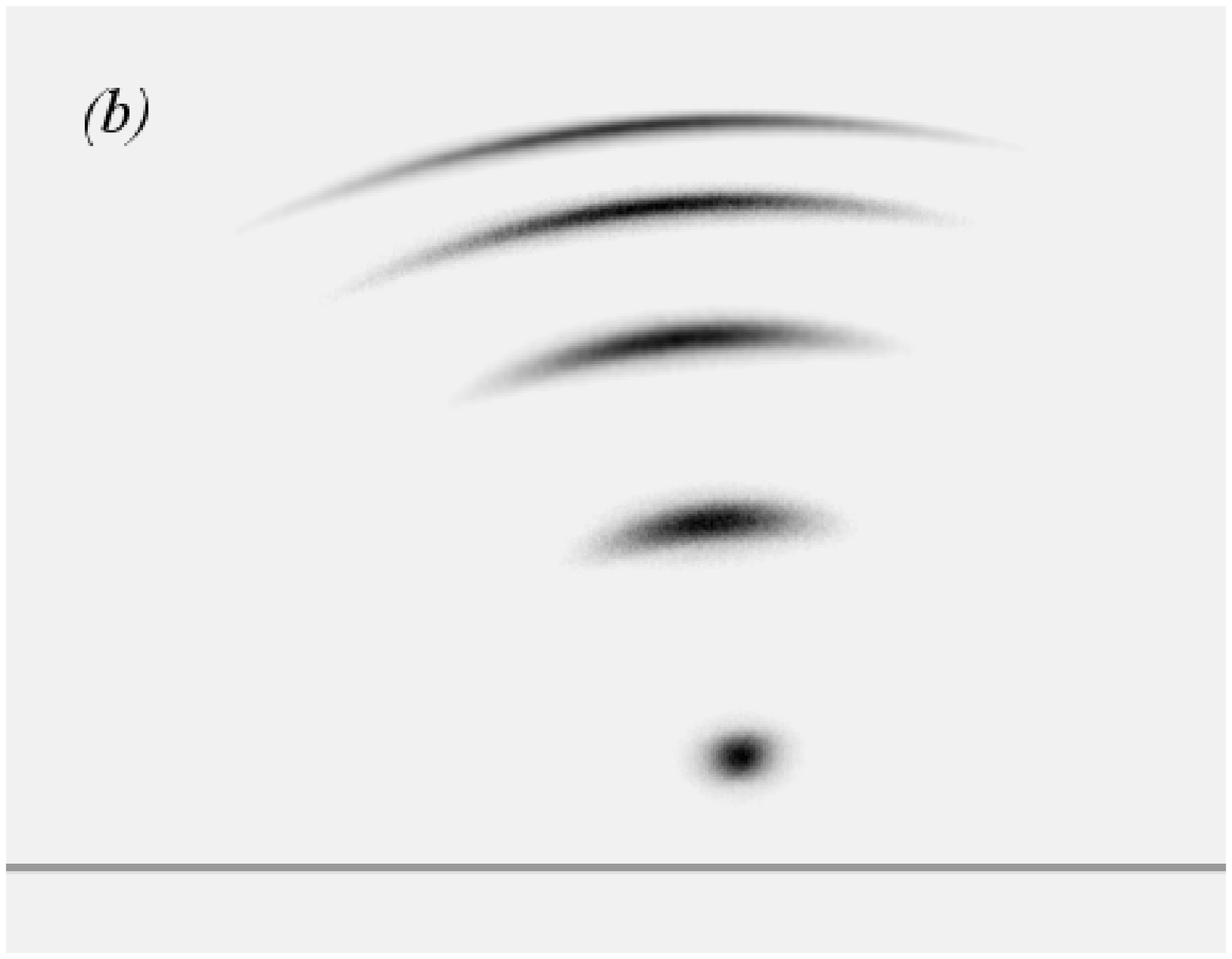}
\end{center}
\caption{\textit{(a)} Absorption imaging pictures of a BEC with $3\times 10^5$ atoms bouncing off a diffuse mirror, for different delays after reflection: 2~ms, 7~ms, 12~ms, 17~ms and 22~ms (the pictures are superimposed). The prism surface, slightly tilted from the imaging axis, is visible at the bottom of the picture. \textit{(b)} Simulation of a bouncing BEC in the same conditions, with a velocity spread along $y$ chosen to be $\sigma_{v_y} = \sigma_{v_x} / 2 = 19.5$~mm.s$^{-1}$. The position of the mirror surface is marked by a grey line. The dimensions of both pictures are 5.7~mm $\times$ 4.4~mm.}
\label{bounce}
\end{figure}

The picture taken after reflection show a strong atomic scattering (figure~\ref{bounce}$a$). The atoms are spread onto a sphere with a radius increasing approximately linearly with time. From the Gaussian $1/\sqrt{e}$ radius of the atomic cloud along $x$, we infer a velocity width of $\sigma_{v_x}=39.4$~mm/s (Gaussian $1/\sqrt{e}$ radius), corresponding to $6.6~v_{\mbox{\scriptsize rec}}$ where $v_{\mbox{\scriptsize rec}} = 5.89$~mm.s$^{-1}$ is the recoil velocity for rubidium. Note that the initial velocity width contributes to less than 1 per cent to this value. This strong atom scattering has no relation with the weak spontaneous photon scattering, responsible for a slight blur of the picture (on the order of 1 pixel rms of the final $512 \times 512$ image).

\subsection{Analysis of $y$ spreading}
It is more difficult to have access to the velocity spreading $\sigma_{v_y}$ along $y$, as it corresponds to the imaging axis. Nevertheless, it can be deduced from a careful analysis of the data, in conjunction with a numerical simulation reproducing the experiment for different possible values of the $y$ velocity spreading. It is clear for instance that, in the case of a totally isotropic scattering, the atomic cloud would shade on the $xz$ plane towards the surface. In contrast, a one dimensional scattering along $x$ would present a circle line.

\subsubsection{Numerical simulation}
The experimental picture is compared with a picture calculated for a diffuse reflection, knowing all initial parameters of the expansion and the amplitude $\sigma_{v_x}$ of the scattering along $x$. The corresponding $y$ amplitude $\sigma_{v_y}$ is given as an input parameter and the output image should correspond at best to the experimental image. $N = 3\times 10^5$ classical atomic trajectories are calculated, with random initial positions and velocities which match the initial Thomas-Fermi position and velocity distribution for 40~\% of atoms in the condensate, the other distributions corresponding to those of a thermal cloud at 285~nK in the harmonic trap with known oscillation frequencies. Only the initial position of the condensed part width along $y$ and $z$ is neglected, because its contribution to the total width is very small already after a few milliseconds of ballistic expansion. In the calculation, the mirror is flat and the $z$ component of the atomic velocity is instantaneously reversed, thus neglecting the exponential character of the real mirror: it doesn't play an important role in the dynamics at a such height, the bouncing time being in the order of the microsecond. The mirror roughness is taken into account by adding a random horizontal velocity to the incoming one, distributed on a Gaussian with a fixed width $\sigma_{v_x}$ along $x$ and a tentative width $\sigma_{v_y}$ along $y$. The vertical velocity of each atom is modified correspondingly to ensure energy conservation. The final calculated image is produced by integrating the atomic density along the imaging direction ($y$ axis).

\subsubsection{Deducing the $y$ spreading}

\begin{figure}[t]
\begin{center}
\includegraphics[width=80mm]{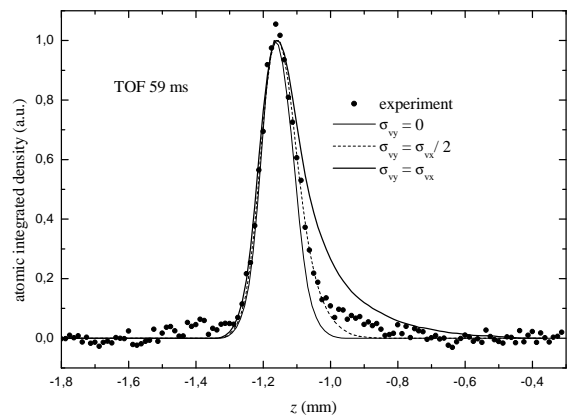}
\caption{Comparison of experimental results and simulation, after 59~ms total time of flight: atomic density profiles integrated along $y$ and averaged over 1.5~mm along $x$. Closed circles: normalized experimental data. Lines: normalized numerical calculation, for different tentative values of the scattering amplitude along $y$ $\sigma_{v_y}$. The parameters for the simulation coincide with the measured experimental parameters: 3.59~mm initial height above the mirror, 40~\% of condensed atoms, Thomas-Fermi radii $V_y = V_z = V_{\perp} = 5.96$~mm.s$^{-1}$, $V_x = 0.89$~mm.s$^{-1}$, $R_x = 90~\mu$m, temperature $T = 285$~nK, initial velocity $v_x = -30.67$~mm.s$^{-1}$, $v_z = 0.3$~mm.s$^{-1}$ and scattering along $x$ $\sigma_{v_x} = 39$~mm.s$^{-1}$. Thin line:  $\sigma_{v_y} = 0$ (1D scattering); dashed line: $\sigma_{v_y} = \sigma_{v_x}/2$ (medium anisotropic scattering); bold line: $\sigma_{v_y} = \sigma_{v_x}$ (isotropic scattering).}
\label{profiles}
\end{center}
\end{figure}

The direct comparison of the experimental and calculated picture confirms that the simulation nicely reproduces the overall behaviour of the atoms (figure~\ref{bounce}). To be more quantitative, we isolate a small region of size 0.8~mm $\times$ 1.5~mm along $x$ and $z$ respectively, centered on the peak density, and compare the profiles deduced after integration along $x$ for both pictures (the integration cannot be performed over the whole sample due to the curvature of the signal). Figure~\ref{profiles} gives an example of such profiles, after a 59~ms total time of flight.

As can be seen on the figure, the experimental results are not consistent with an isotropic scattering, neither with a fully 1D scattering. A reasonable value for $\sigma_{v_y}$ would be $\sigma_{v_x}/2 = 3.3~v_{\mbox{\scriptsize rec}}$, that is an anisotropy $\chi = \sigma_{v_y}/\sigma_{v_x} = 2$. We cannot exclude values of $\chi$ as low as 1.5 or as high as 2.5; in any case, the scattering is larger in the direction of propagation of the real part $K_{\rm ev} \mathbf{e_x}$ of the optical wave vector. 

\section{Theory}
\label{theory}

\subsection{Presentation of theory}
The problem of diffuse reflection of atoms off an evanescent wave mirror has been addressed theoretically in Ref~\cite{Henkel97a} and it linked to the scattering of light by the defects of a rough dielectric surface. Atom scattering is due to Raman stimulated transitions involving a photon of the evanescent mode and a scattered photon. The expected momentum distribution after reflection can be calculated from the knowledge of the roughness power spectrum $P_S( {\bf Q} )$ of the surface and an atomic response function $B_{\rm at}( {\bf Q} )$ [Eqs.(5.6, 5.7) of 
Ref.\cite{Henkel97a}]. $\bf Q$ is an in plane two dimensional wave vector. For normal incidence onto the surface, the variance of the velocity distribution after the bounce is then given by
\begin{equation}
\frac{ \sigma_{v_i}^2 }{ v_{\rm rec}^2 } = \frac{\mathcal{N}}{k_L^2} \int\!{\rm d}^2Q \,
Q_i^2 P_S( {\bf Q} ) |B_{\rm at}( {\bf Q} )|^2 , \quad \quad i=x,y.
\label{eq:dQx-theory}
\end{equation}
$\mathcal{N}$ is a normalization factor, related to the scattering probability $w$ through
\begin{equation}
\mathcal{N}=w/ \int\!{\rm d}^2Q \, P_S( {\bf Q} ) |B_{\rm at}( {\bf Q} )|^2
\end{equation}
Equation (\ref{eq:dQx-theory}) allows one to compute the anisotropy of the velocity width after reflection, the dependence on the angle of incidence being actually negligible for our parameters (see Ref.\cite{Henkel97a}).

The roughness power spectrum may be inferred from a measurement of the surface. For most surfaces, this spectrum follows a power law $P_S( {\bf Q} ) \propto 1/Q^{\alpha}$, with $\alpha$ being typically between 2 and 5. The atomic response function $B_{\rm at}( {\bf Q} )$ has a complicated expression. Nevertheless, for atomic incidence normal to the surface, it is very peaked around two circles centered at $\pm K_{\rm ev} \mathbf{e_x}$ of radius $k_L$. We will thus model this function by a delta function around these circles, $|B_{\rm at}( {\bf Q} )|^2 \propto \delta(|\mathbf{Q} - K_{\rm ev} \mathbf{e_x}| - k_L) + \delta(|\mathbf{Q} + K_{\rm ev} \mathbf{e_x}| - k_L)$. Note that this corresponds to the absorption of a photon from the guided mode and the successive stimulated emission into a free mode at grazing incidence -- or vice-versa. These two processes contributes to the same amount to $\sigma_{v_i}^2$, as $P_S( Q )$ only depends on the modulus of the transferred wave vector. The integration may be performed around the value $\mathbf{Q} = K_{\rm ev} \mathbf{e_x}$, leading to:
\begin{equation}
\frac{ \sigma_{v_x}^2 }{ v_{\rm rec}^2 } \simeq \frac{2\mathcal{N}}{k_L^2} \int\!{\rm d}^2Q \,
(Q_x - K_{\rm ev})^2 P_S( \textbf{Q} - K_{\rm ev} \mathbf{e_x} ) \delta(Q - k_L)
\end{equation}
\begin{equation}
\frac{ \sigma_{v_y}^2 }{ v_{\rm rec}^2 } \simeq \frac{\mathcal{2N}}{k_L^2} \int\!{\rm d}^2Q \,
Q_y^2 P_S( \textbf{Q} - K_{\rm ev} \mathbf{e_x} ) \delta(Q - k_L)
\end{equation}
After integration and simplification, with a power law for the roughness spectrum, we get:
\begin{equation}
\chi^2 = \frac{ \sigma_{v_x}^2 }{ \sigma_{v_y}^2 } = \frac{ \displaystyle \int_0^{\pi} d \varphi \frac{( \cos \varphi - \eta )^2}{(1 - 2 \eta \cos \varphi + \eta^2)^{\alpha/2}} }{ \displaystyle \int_0^{\pi} d \varphi \frac{\sin^2 \varphi}{(1 - 2 \eta \cos \varphi + \eta^2)^{\alpha/2}} }
\end{equation}
where $\eta = K_{\rm ev} / k_L = n_g \sin \theta_i$. This expression depends only slightly on the choice of $\alpha$: $\chi$ lies between 2.05 and 2.31 for values of $\alpha$ between 2 and 5 which are relevant for a typical polished dielectric surface. It is minimum for $\alpha = 3$ and has the same value for $\alpha$ and $6-\alpha$. In the case $\alpha = 4$, the expression for $\chi^2$ is very simple as we have $\chi^2 = 1 - 2 \eta^2$, which for our parameters leads to $\chi = 2.12$, in good agreement with the experimental data.

\subsection{Estimation of momentum spreading}
The evaluation of the expected momentum spreading is more delicate, as a full knowledge of $B_{\rm at}( {\bf Q} )$ and $P_S( {\bf Q} )$ would be required to estimate $w$. However, if one knows the rms surface roughness, it is possible to give an upper bound of the effect using the upper limit for the total diffuse reflection probability $w_{\rm max}$ (Eq. (5.21) of Ref.~\cite{Henkel97a}):
\begin{equation}
w < w_{\rm max} = \left( 4 \pi \sigma_s e^{\kappa z_0}/\lambda_{\rm dB} \right)^2
\end{equation}
Here $\lambda_{\rm dB}=h/mv_z$ is the mean de Broglie wave length of the incoming cloud, $z_0$ is the height of the classical reflection turning point, and $\sigma_s$ is the surface roughness defined as the square root of the variance of the surface height. The velocity spreading may then be estimated as being, for example in the $x$ direction:
\begin{equation}
\sigma_{v_x}^2 = v_{\rm rec}^2 w_{\rm max} \frac{ \displaystyle \int_0^{\pi} d \varphi \frac{( \cos \varphi - \eta )^2}{(1 - 2 \eta \cos \varphi + \eta^2)^{\alpha/2}} }{ \displaystyle \int_0^{\pi} \frac{d \varphi}{(1 - 2 \eta \cos \varphi + \eta^2)^{\alpha/2}} }
\end{equation}
A surface roughness of $\sigma_s=3.3$~nm was deduced from an AFM measurement of a portion of the prism. For $\alpha=4$, and a falling height of 3.6~mm, we find an expected maximum velocity spreading along $x$ of $8.00~v_{\rm rec}$, a bit higher than the observed $6.6~v_{\rm rec}$.

\section{Conclusion}
In conclusion, we have performed measurements on the diffuse reflection of a Bose-Einstein condensate on a rough surface. The experiment shows a strong scattering of the atomic cloud, due to elastic Raman photon scattering involving both the mirror light and scattered light. A first clear evidence of the anisotropy of this diffuse reflection is given~\cite{Chris}. The scattering is twice along the direction of propagation of the evanescent mode, and this value is in good agreement with the prediction of Henkel \textit{et al.}~\cite{Henkel97a}, in a simplified model. A tentative absolute value for the scattering amplitude was also given, in a reasonable agreement with the experiment. However, a more precise analysis would be necessary to confirm the anisotropy value and the absolute value~\cite{Colombe05a}.

The experiments were performed with a Bose-Einstein condensate. However, the coherence of the source is not visible on the results -- no speckle is visible for instance. An estimation of the size of the speckle spots, on the order of $\lambda_{\rm dB} h_0/R$ where $h_0=3.6$~mm is the distance between cloud and surface and $R\sim100\mu$m its size at the reflection point, would give typically 1~$\mu$m, which is not resolved by the imaging system. Nevertheless, the reflection should be coherent, the spontaneous photon scattering being negligible. To prevent diffuse reflection, a great care should be taken in the surface quality. We point out however that the atom scattering did not mask the diffraction pattern in a temporal diffraction experiment we performed with the very same prism~\cite{Colombe05b}. Interferometry with evanescent wave devices remains a very promising alternative.

\begin{acknowledgments}
We thank C.~Westbrook for stimulating discussions, and for communicating us unpublished data. We gratefully acknowledge support by the R\'egion Ile-de-France (contract number E1213) and by the European Community through the Research Training Network ``FASTNet'' under contract No. HPRN-CT-2002-00304 and Marie Curie Research Network ``Atom Chips'' under contract No. MRTN-CT-2003-505032. Laboratoire de physique des lasers is UMR 7538 of CNRS and Paris 13 University.
\end{acknowledgments}

\end{document}